\documentclass[prl,twocolumn,aps,superscriptaddress,showpacs,floatfix]{revtex4}
\usepackage{longtable}
\usepackage{graphicx}
\usepackage{dcolumn}
\usepackage{bm}
\usepackage{mathbbol}
\usepackage{color}
\newcommand{\vf}{v_{\mbox{\tiny F}}}
\begin{document}
\title{Current resonances in graphene with time dependent potential barriers}
\author{Sergey E. Savel'ev}
\affiliation{Department of Physics, Loughborough University,
Loughborough LE11 3TU, United Kingdom}
\author{Wolfgang H\"ausler}
\affiliation{Institut f{\"u}r Physik,
Universit{\"a}t Augsburg, D-86135 Augsburg, Germany}
\author{Peter H\"anggi}
\affiliation{Institut f{\"u}r Physik,
Universit{\"a}t Augsburg, D-86135 Augsburg, Germany}
\begin{abstract}
A method is derived to solve the massless Dirac-Weyl equation describing
electron transport in a mono-layer of graphene with a scalar potential barrier
$U(x,t)$, homogeneous in the $y$-direction, of arbitrary $x$-
and time dependence. Resonant enhancement of both electron backscattering
and currents, across and along the barrier, is predicted when the modulation
frequencies satisfy certain resonance conditions. These
conditions resemble those for Shapiro-steps of driven Josephson
junctions. Surprisingly, we find a non-zero $y$-component of
the current for carriers of zero momentum along the $y$-axis.
\end{abstract}
\maketitle

Growing interest to graphene, see e.g.\ Ref.\
[\onlinecite{reviews}], is stimulated by many unusual and
sometimes counter intuitive properties of this two dimensional
material. Indeed, graphene supplies charge carriers exhibiting
pseudo-relativistic dynamics of massless Dirac fermions. As
one consequence the Klein tunneling phenomenon \cite{paradox}
occurs with unit probability through arbitrarily high and thick
barriers at perpendicular incidence, irrespective of the
particle energy, in accordance with experiment
\cite{paradox-exp}. The question arose how to control the
electron motion in graphene and hence boosted detailed studies
of Dirac fermions under the influence of various forms of scalar
\cite{superl,super5,left,peeters09b,pnexperiments} or vector
\cite{magstruc} potentials.

Applying a time-dependent laser field to pristine graphene opens
an alternative and efficient way \cite{Efetov,laser1,laser2} to
control spectrum and transport properties. It was shown
\cite{Efetov} that Dirac fermions accross $p$-$n$-junctions can
aquire an effective mass when driven by a laser field. This results
in an exponential suppression
of chiral tunneling even for perpendicular incidence upon the
junction, if the {\sc ac}-electric field is directed parallel to
the junction, in stark contrast to Klein tunneling occuring in
the absence of the laser field. Actually, time dependent laser
fields can mimic \cite{laser2} the influence of any
electrostatic graphene superlattices on the electron spectrum in
graphene. The question arises whether and under which
conditions time-dependent modulations of an electrostatic
barrier, where energy is not conserved, would affect electron
transport and generate backscattering.

In this Letter we answer this question by solving the problem for
arbitrary space-time dependent scalar potentials $U(x,t)$. Our
solution is based on expanding the wave function as a power
series with respect to the momentum $k_y$ parallel to the
barrier, and manifests a structure of left and right moving waves.
All terms appearing in the $k_y$-expansion can be
calculated analytically, despite of the fact that each term is
described by a partial differential equation in $(x,t)$-space. At
$k_y=0$ (normal incidence upon the barrier) we confirm complete
Klein tunneling for any $U(x,t)$ while for finite $k_y$
backscattering resonances can occur at certain angles of
incidence, depending on the modulation frequency of the barrier. 
As a counter intuitive result we find a
non-zero and oscillating current $j_y$ {\it along} the barrier,
even at $k_y=0$ for valley polarized fermions. At $k_y\ne 0$
the current $j_y$ arises also in valley unpolarized situations,
it can be resonantly amplified and flow in either direction.
Interestingly, $j_y$ exhibits a non-zero {\sc dc}-component at
certain resonance frequencies, in full analogy to Shapiro-steps
of driven Josephson junctions.

At low energies, the honeycomb lattice of graphene engenders two copies,
$\tau_z=\pm 1$, of Dirac-Weyl Hamiltonians \cite{kanemele}
\begin{equation}\label{h0}
H_0=\vf\:[\hat\tau_z\hat\sigma_x\hat p_x+\hat\sigma_y\hat p_y]\;,
\end{equation}
centered about two inequivalent Dirac points (``valleys'') $K$
and $K'$ at corners of the hexagonal first Brillouin zone where
electron-hole symmetric bands touch; Pauli matrices
$\hat\sigma_{x,y,z}$ act on two-component spinors representing
sublattice amplitudes. Carriers near either of the Dirac points
exhibit opposite Fermion helicities,
$\:\bm{\sigma}\cdot\bm{p}/p=\pm 1\:$. Proposals exist in
literature how to valley polarize carriers in graphene, by means of
nanoribbons terminated by zig-zag edges \cite{zigzagribbons}, by
exploiting trigonal warping at elevated energies
\cite{garcia08}, or by absorbing magnetic textures
\cite{ziegler10}.

Smooth electromagnetic or disorder potentials
\cite{ando98}, containing negligible fourier components at large
wave vectors of the order of $|{\vec K}|$, will not cause scattering between valleys so that calculations 
can be carried out for $\tau_z=+1$ or $\tau_z=-1$ separately. Accordingly,
time dependent potentials $U(x,t)$ should be slowly varying,
without frequency components that might induce excursions to
energies where the band structure of graphene starts deviating
from the isotropic cone spectrum, i.e. below 0.6\,eV
\cite{zhou06}. Including $U(x,t)$, the Dirac equation for the wave
function $\Psi_{k_y}(x,y,t)= \Psi(x,t)\exp(ik_yy)$ can be
written in the form
\begin{equation}\label{sup}
\left(\begin{array}{cc}U(x,t)& -i\tau_z\frac{\partial}{\partial x}\\
-i\tau_z\frac{\partial}{ \partial x} & U(x,t)\end{array}\right)
\Psi+ik_y\left(\begin{array}{cc}0& -1\\ 1& 0\end{array}\right)
\Psi=i\frac{\partial}{ \partial t}\Psi
\end{equation}
where from now on we assume $\vf=1$ and $\hbar=1$. This equation
has been solved analytically for time-independent potentials
either by matching \cite{paradox} of wave functions for
rectangular barriers, or by the WKB method \cite{wkb-method} for
smooth barriers. Time dependent harmonic oscillations have been
considered of gate voltages on either side of a graphene
rectangle \cite{harmonicgates}, of an electric field parallel to
the barrier \cite{Efetov} or in resonance approximation
\cite{laser2}, or for some class of time dependent barriers
$U(x,t)$ at $k_y=0$ \cite{solomon}.

Our goal here is to construct the solution of eq.~(\ref{sup})
for arbitrary $U(x,t)$ acting at positive times, $\:U(x,t<0)=0\:$.
From the Ansatz
\begin{equation}\label{series}
\Psi=\sum_{n=0}^{\infty}(ik_y)^n\left(\begin{array}{c}1\\ \tau_z\end{array}
\right)\Psi_{n,+} +\sum_{n=0}^{\infty}(ik_y)^n
\left(\begin{array}{c}1\\ -\tau_z \end{array} \right)\Psi_{n,-}
\end{equation}
as a power series in $k_y$ we derive a recurrence relation for
the coefficients $\Psi_{n,\pm}$ which obey the inhomogeneous
first order partial differential equations,
\begin{equation}\label{set1}
\left(U(x,t)\mp i\frac{\partial}{\partial x}-i\frac{\partial}{\partial
t}\right)\Psi_{n,\pm}\pm\tau_z\Psi_{n-1,\mp}=0\;,
\end{equation}
with $\Psi_{-1,\pm}(x,t\geq 0)\equiv 0$. Initial conditions can be
chosen as
$\Psi_{0,\pm}(x,t=0)=a_{\pm}(x)=[\Psi_A(x,t=0)\pm\tau_z\Psi_B(x,t=0)]/2$,
$\Psi_{n>0,\pm}=0$, where $\Psi_A, \Psi_B$ describe electron
amplitudes on either of the graphene sublattices. The two
functions $a_{\pm}(x)$, providing the initial conditions,
can be, e.g., a plane wave or a wave packet. We underline here the
general structure of (\ref{series}) as a sum of right $\Psi_+$
and left $\Psi_-$ moving waves.
Using the standard d'Alembert's ratio test, a sufficient
criterion for convergence of series (\ref{series}) is
$|k_y|\lim_{n\rightarrow\infty}|\Psi_{n+1,\pm}|/|\Psi_{n,\mp}|<1$
for all relevant $x$ and $t$.

Despite of the fact that (\ref{set1}) are partial differential
equations, we can solve them exactly using the method of
characteristics \cite{couranthilbert}. The corresponding result reads
\begin{equation}\label{coeffs}
\Psi_{n,\pm}(x,t)=a_{n,\pm}(x,t)e^{-i\int_0^tdt'\;U(x\mp t\pm t', t')}
\end{equation}
with $a_{0,\pm}=a_{\pm}(x\mp t)$ and
\begin{eqnarray*}
a_{n>0,\pm}&=&\mp i\tau_z\int_0^{t}dt'\;\Psi_{n-1,\mp}(x\mp t\pm t',t')\\
&&\times e^{i\int_0^{t'}dt''\;U(x\mp t\pm t'',t'')}\;.
\end{eqnarray*}

Together with (\ref{series}) the recursive solution for
$\Psi_{n,\pm}$ provides the exact wave function $\Psi$ to any
desired accurancy.
To zeroth order approximation w.r.t.\ $k_y$ we obtain:
\begin{eqnarray}
&&\psi(x,t)=a_+(x-t)\left(1\atop \tau_z \right)
e^{-i\int_0^tdt'\; U(x-t+t',t')}
\nonumber \\ &&\;\mbox{}+a_-(x+t)\left(1\atop -\tau_z\right)
e^{-i\int_0^tdt'\; U(x + t-t',t')}\;.
\label{timeevolution0}
\end{eqnarray}
The first order corrections w.r.t.\ $k_y$ in
(\ref{coeffs}) can be written as
\begin{eqnarray}
a_{1,\pm}=\mp i\tau_zA_{1,\pm}=\mp i\tau_z\int_0^tdt'\;
a_{\mp}(x\mp t\pm 2t')\nonumber \\
\times e^{i\int_0^{t'}dt''\;[U(x\mp t\pm t'',t'')-U(x\mp t\pm 2t'\mp
t'',t'')]}\;,
\label{a-one}
\end{eqnarray}
so that $\Psi=\Psi_+\left({1\atop\tau_z}\right)+
\Psi_-\left({1\atop -\tau_z}\right)$ as in (\ref{series}), with
\begin{equation}\label{psi-pm}
\Psi_{\pm}=[a_{0,\pm}\pm k_y\tau_zA_{1\pm}]e^{-i\int_0^tdt'\:U(x\mp
t\pm t',t')}\;.
\end{equation}
When $k_y=0$ and when the wave packet is initially purely right
moving, $\:a_-=0\:$, eq.~(\ref{timeevolution0}) reveals that the electron
density distribution $\:|a_+(x-t)|^2\:$ undistortedly continues
to propagate to the right without reflection: $\Psi_-(x,t)=0$ for
all times $t>0$. This proves complete Klein tunneling also in
the presence of time dependent barriers; wave functions
$\Psi_{\pm}$ acquire only a phase factor by the potential
at $k_y=0$.

As a measurable quantity, we now evaluate the current density in
cartesian components, $j_x=\Psi^*\tau_z\sigma_x\Psi=2\Psi_{+}^*
\Psi_+-2\Psi_{-}^*\Psi_-=j_{0x}+j_{1x}$ and
$j_y=\Psi^*\sigma_y\Psi=2i\tau_z(\Psi_{+}^*\Psi_--\Psi_+\Psi_{-}^*)=
j_{0y}+j_{1y}$. Here, the last equal signs refer to zeroth and
first order contributions w.r.t.\ $k_y$, respectively, yielding
\begin{equation}\label{jx0}
j_{0x}=2(|a_{0+}|^2-|a_{0-}|^2)
\end{equation}
\begin{equation}\label{jy0}
j_{0y}=4\tau_z|a_{0+}a_{0-}|\sin(\varphi+\phi_0)
\end{equation}
\begin{equation}\label{jx1}
j_{1x}=4k_y\tau_z\Re{\rm e}\{a_{0+}A_{1+}^*-a_{0-}A_{1-}^*\}
\end{equation}
\begin{equation}\label{jy1}
j_{1y}=4k_y\left(|A_{1+}a_{0-}^*|\sin(\varphi-\phi_-)-
|A_{1-}^*a_{0+}|\sin(\varphi+\phi_+)\right)
\end{equation}
with $\varphi=\int_0^t[U(x+t-t',t')-U(x-t+t',t')]dt'$,
$\phi_0={\rm arg}(a_{0+}a_{0-}^*)$, and $\phi_{\pm}={\rm
arg}(a_{0\pm}A_{1\mp}^*)$. We distinguish two cases: (i)
$\tau_z$-independent contributions $j_{0x}$ and $j_{1y}$ which
can be observed for valley unpolarized carriers and (ii)
$\tau_z$-dependent contributions $j_{1x}$ and $j_{0y}$ where
detection calls for valley polarization.

Eqs.~(\ref{jx0}) and (\ref{jy0}) describe the current density at
normal incidence, $k_y=0$. Then $j_x$ stays unaffected by the
barrier, irrespective of $\tau_z$ which rephrases the above
result of complete Klein tunneling. Surprisingly, a current $j_y$
flows perpendicular to the momentum in graphene, provided the
sample is valley polarized (the total current $\bm{j}=\bm{j}^+
+\bm{j}^-$, where $\bm{j}^{\tau_z}$ originates from states near
valley $K$ ($\tau_z=+1$) or $K'$ ($\tau_z=-1$), respectively).
This current (\ref{jy0}) results from interfering left and right
moving waves, which both need to have nonzero amplitudes,
$a_{0+}a_{0-}\ne 0$. 

Eqs.~(\ref{jx1}) and (\ref{jy1}) describe corrections to the
current density at small but finite angles of incidence, $k_y\ne
0$. Thereby, $j_{1x}$ exhibits qualitatively similar properties as
$j_{0y}$; in particular it stays nonvanishing at finite valley
polarization only. By contrast, the current density $j_{1y}$ 
now exhibits striking current oscillations and
current reversals already in valley unpolarized situations as we
show in more detail below.

Next we turn to the question how carriers are reflected by
$U(x,t)$. Let's consider an initially right moving plane wave,
$\Psi_{k_y}=e^{i(k_xx+k_yy)}\left({1\atop\tau_z}\right)$
at $t=0$ which produces a current
density $j_{0x}=+2$ pointing to the right. Using equations
(\ref{series}) and (\ref{psi-pm}), and assuming small $k_y$, the
leading contribution to the reflected current density
$j_{2x}=-k_{y}^2|A_{1-}|^2$ arises in ${\cal O}(k_y^2)$ under
the action of the barrier at $t>0$ and is proportional to
$|j_{0x}|$, cf.\ (\ref{a-one},\ref{jx0}). This suggests
to employ the ratio
\begin{equation}\label{reflect}
R(x,t):=-j_{2x}/j_{0x}=k_{y}^2|A_{1-}(x,t)|^2/2
\end{equation}
as a measure for the reflectivity at small $k_y$. While the
quantity $R(x,t)$ evolves in time, together with $U(x,t)$, it is
independend of $\tau_z$ and, thus, measurable without valley
polarization. Moreover, we also analyze the time averaged
reflectivity $\:\overline{R}(x):=\:\lim_{T\rightarrow
\infty}\int_0^{T}R(x,t)dt/T$, which can be measured just by
means of {\sc dc}-equipment.

In the following, two specific examples $U(x,t)=U^{(1,2)}(x,t)$
are considered. As initial conditions we take into account two
cases: (i) a superposition of equal amplitudes of right and left
propagating plane waves, $a_{\pm}=\exp(\pm ik_xx)$ and
$\tau_z=+1$ when calculating $j_{0y}$; (ii) an incidently right
moving wave $a_+=\exp(ik_xx), a_-=0$ when calculating $R$ and
$j_{1y}$ for valley unpolarized systems. Our first example is
$\:U^{(1)}(x,t)=U_0x\sin\omega t\:$. In view of (\ref{jy0}), we
derive for this case
\begin{equation}\label{jy11}
\textstyle j_{0y}(x,t)=\:\sin
2\left[k_xx+U_0\left(\frac{t}{\omega}-\frac{\sin\omega
t}{\omega^2}\right)\right]\;,
\end{equation}
which can be rewritten as a sum
\begin{equation}\label{jy11n}
j_{0y}(x,t)=\sum_{n=-\infty}^{\infty}J_n\left(\frac{2U_0}{\omega^2}\right)
\sin\left(2k_xx+\frac{2U_0t}{\omega}-n\omega t\right)\;,
\end{equation}
using Bessel functions $J_n$. This form (\ref{jy11n}) reveals
a peculiarity at $\omega=\omega_n$ with
\begin{equation}\label{resomn}
\omega_n=\sqrt{2U_0/n}\;,\quad n\in\mathbb{N},
\end{equation}
similar to Shapiro-steps \cite{tinkham} of a driven Josephson
junction. As depicted in Fig.~1a, frequencies $\omega=\omega_n$
generate periodic oscillations, which, again as in the case of
Shapiro-steps, induce a nonzero {\sc dc}-component in the
current at given $x$. Modulating the potential with
$\omega\ne\omega_n$ results in aperiodic oscillations and zero
{\sc dc}-component.

\begin{figure}
\includegraphics[width=9cm]{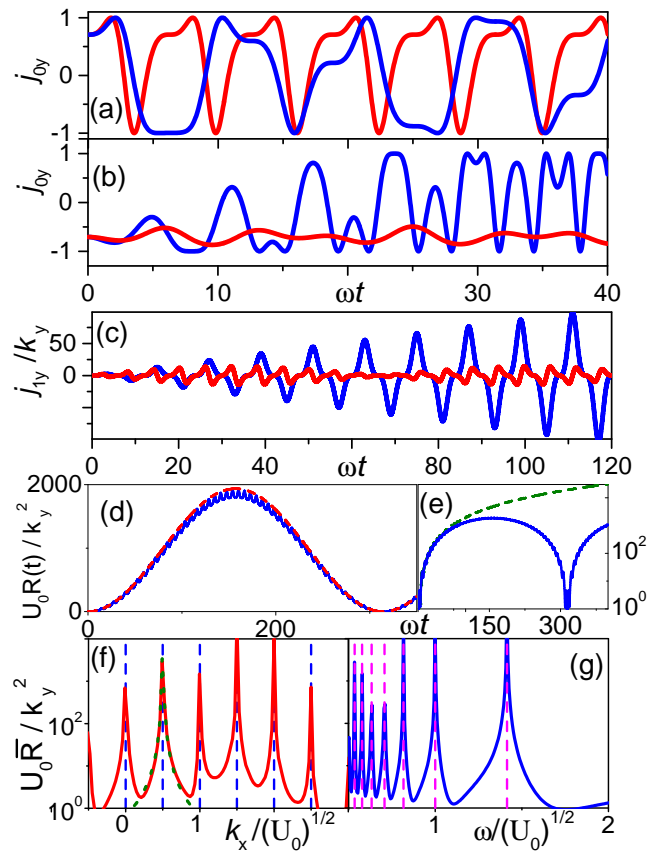}
\caption{(Color online) (a) Current $j_{0y}$ (\ref{jy0})
perpendicular to $\bm{k}$ versus time for
$U^{(1)}(x,t)=U_0x\sin\omega t$, $k_xx=\pi/8$, and
$U_0/\omega^2=1/\pi$ (blue line) and $U_0/\omega^2=1/2$ (red
line), assuming a valley-polarized situation $\tau_z=1$. For
``Shapiro-step'' conditions (eq.~(\ref{resomn})) periodic
oscillations can be seen (red), while, away from this condition,
aperiodic oscillations occur (blue).
(b) Same as (a) but for potential
$U^{(2)}(x,t)=U_0\cos(x/L)\cos\omega t$ with $x=\pi L/2,\ k=0,\
U_0L=0.1$, and frequencies $\omega=(\pi/2)(1/L)$ (red line) and
$\omega=1/L$ (blue line). Both currents are aperiodic. For the
matching condition $\omega=1/L$ (blue) a considerable
enhancement followed by a saturation of the amplitude of the
current oscillations occurs, while away from this resonance no
enhancement is seen versus time.
(c) Current $j_{1y}$ (\ref{jy1}) for $U^{(1)}(x,t)$ at $x=0$,
using $k=0$. At Shapiro-resonance ($U_0/\omega^2=1/2$, blue
line) pronounced current enhancement occurs, cf.\
eqs.~(\ref{a1minus},\ref{angleresonance}) while away from the
resonance ($U_0/\omega^2=3/\pi$, red line) no enhancement is
seen.
(d) Time-dependent reflectivity $R(t)$, calculated by numerical
integration of eq.~(\ref{a1minus}), blue line, and by using the
approximation (\ref{rt}), red dashed line, near resonance for
$k_x/\sqrt{U_0}=0$ and $\omega^2/U_0=0.49$.
(e) $R(t)$ at resonance (green dashed line, $k_x/\sqrt{U_0}=0,
\omega^2/U_0=1/2$) and near resonance (blue line, $k_x/\sqrt{U_0}=0,
\omega^2/U_0=0.49$). At the resonance, $R(t)$ increases with
time $\sim t^2$, in agreement with equation (\ref{rt}).
(f) Time-averaged reflectivity $\overline{R}$ as a function of
$k_x/\sqrt{U_0}$ for $\omega/\sqrt{U_0}=1$. Equidistant
resonances occur at $k_x/\sqrt{U_0}=1-n/2$ (dashed blue vertical
lines), cf.\ eq.~(\ref{angleresonance}). One of
the resonance peaks is well fitted by the resonance equation
(\ref{av-r}), as shown in dashed green.
(g) $\overline{R}$ as a function of the driving frequency
$\omega/\sqrt{U_0}$ for $k_x/\sqrt{U_0}=0$: resonance peaks are
clearly seen at Shapiro-step conditions (\ref{angleresonance})
$\omega/\sqrt{U_0}=\sqrt{2/n}$, indicated as dashed magenta
vertical lines.}
\label{fig1}
\end{figure}

Similar resonance effects can also be seen in both, the
reflectivity $R$ (\ref{reflect}) and the current $j_{1y}$
(\ref{jy1}). By inserting $U^{(1)}(x,t)$ into eq.~(\ref{a-one})
we derive
\begin{eqnarray}\label{a1minus}
A_{1-}&=&e^{ik_x(x+t)}\int_0^tdt'\: e^{-2ik_xt'}
e^{2iU_0(\omega t'-\sin(\omega t'))/\omega^2}\\
&=&e^{ik_x(x+t)}\sum_{n=-\infty}^{\infty} J_n\left(\frac{2U_0}{\omega^2}\right)
\frac{e^{i(2U_0/\omega-2k_x-n\omega)t}-1}{i(2U_0/\omega-2k_x-n\omega)}
\;,\nonumber
\end{eqnarray} 
from which we read off a Shapiro-step resonance condition
\begin{equation}\label{angleresonance}
k_n=k_x=-\frac{n}{2}\omega+\frac{U_0}{\omega}\;, \quad n\in\mathbb{N}
\end{equation}
specifying now a {\it directional\/} dependence of the momentum
$\bm{k}$. From (\ref{a1minus}) together with (\ref{jy1}) we
conclude that in valley-unpolarized samples the current
$j_y\propto k_y$ parallel to the barrier oscillates as a
function of time and may take either sign (despite of the fixed
$k_y$, see Fig.~\ref{fig1}c). In addition, the amplitude of
these oscillations increases with time as the Shapiro-step
resonance condition (\ref{angleresonance}) is met (compare red
and blue curves in Fig.~\ref{fig1}c).

Analogous resonances also show up in both reflectivities, $R$
and $\overline{R}$. The latter allows experimental observation
of the here predicted behavior without time-domain measurements.
Indeed, near the Shapiro-step resonance (\ref{angleresonance})
we can keep only one summand in the expansion (\ref{a1minus}),
yielding
\begin{equation}\label{rt}
R(t)=\frac{1}{2}k_{y}^{2}J_{n}^{2}\left(\frac{2U_0}{\omega^2}\right)
\frac{\sin^2[(U_0-k_x-n\omega/2)t]}{(U_0/\omega-k_x-n\omega/2)^2}\;.
\end{equation}
This equation is in a good agreement with numerical integration
of (\ref{a1minus}), see Figs.~\ref{fig1}d,e. Averaging
(\ref{rt}) with respect to time results in
\begin{equation}\label{av-r}
\overline{R}=\frac{1}{4}k_{y}^{2}J_{n}^{2}\left(\frac{2U_0}{\omega^2}\right)
\frac{1}{(U_0/\omega-k_x-n\omega/2)^2}\;,
\end{equation}
so that the barrier will become intransparent near momenta
$k_x=k_n$ (see Fig.~\ref{fig1}f), already for small $U_0$. This
produces strong anomalies in transport properties at angles
$\arctan(k_n/k_y)$ of the incidence. Instead of sweeping the
directions of $\bm{k}$ one may alternatively sweep $\omega$ at
fixed $\bm{k}$, cf.~(\ref{angleresonance}); ensuing resonance
peaks are clearly observed in Fig.~\ref{fig1}g.
The constraint $R<1$ determins the maximum value
\begin{equation}
|k_y|\lesssim\frac{|U_0-k_x-n\omega/2|}{|J_n(2U_0/\omega^2)|}\;,
\end{equation}
where second and higher order terms in the expansion (\ref{series}) 
can be ignored.

As a second example, we consider $\:U^{(2)}(x,t)=U_0\cos(x/L)\:
\cos\omega t\:$ to demonstrate how even more intriguing
resonance features can arise from the interplay between spatial
{\it and\/} temporal periodicities. Given again the initial
condition of left and right moving plane waves of equal
amplitudes, and assuming valley polarization we find
\begin{eqnarray}\label{u2j0y}
j_{0y}&=&\sin\Bigl\{2kx-\frac{4U_0L\sin(\frac{x}{L})}{\omega^2L^2-1}\\
&\times&\sin\left[\frac{\omega L+1}{2L}t\right]\:
\sin\left[\frac{\omega L-1}{2L}t\right]\Bigr\}\;.\nonumber
\end{eqnarray}
Now, oscillations of $j_{0y}$ persist even when
$\omega\rightarrow 0$, since the spatial periodicity $2\pi L$ of
the potential induces a frequency component $\vf/L$ to waves
moving at the uniform Fermi velocity (restoring here $\vf$).
This reminds of the {\sc ac}-Josephson effect \cite{tinkham}
where {\sc ac}-current oscillations are generated by a
time-independent voltage.

On the other hand, if the barrier modulation frequency
$\omega\rightarrow\pm\vf/L$, the argument of the sine in the curly
brackets (\ref{u2j0y}) varies proportional to $t$ as $\:2k_xx\mp
t\,U_0\sin(x/L)\sin\omega t\:$. For small $U_0$ the oscillations
of $j_{0y}$ thus grow resonantly with time, before they saturate
at $t\gtrsim 2\pi/U_0$, cf.\ Fig.~\ref{fig1}b. We mention the
analogy to resonant excitations of plasmonic oscillations by
spatio-temporal mode matching of the incident light with the
grating period (Wood's anomaly \cite{woods}). Similar effects
occur also for valley unpolarized currents (e.g., $j_{1y}$) and
the reflectivity $R(x,t)$, but calculations become considerably more cumbersome
and will be published elsewhere.

Concluding, we present the analytical solution of the Dirac
equation for Fermions in graphene moving in a scalar potential
barrier $U(x,t)$ of arbitrary $x$- and time-dependence. Unit
transmission probability, referred to as Klein tunneling, is
found for normal incidence upon the barrier, rendering at most a
phase to the wave function. On the other hand, under certain
angles with respect to the barrier ($k_y\ne 0$), we predict
strong reflection, even for weak potentials. Further, also the
current parallel to the barrier, $j_y$, may exhibit
oscillations, despite of a constant electron momentum $k_y$.
The amplitude of these oscillations grows linearly in time when
$U(x,t)$ meets certain resonance frequencies. In
valley-polarized samples $j_y$ does not vanish even for zero
momentum parallel to the barrier ($k_y=0$), provided left and
right moving waves both interfere with finite amplitudes. For
graphene nanostructures driven by oscillating potentials, the
predicted resonances in current and reflectivity can be seen,
for example, in electron transport properties 
(e.g., in AC and DC electrical conductivity) through suitably
arranged quantum point contacts. The new non-stationary
phenomena in graphene calculated here within the single-particle
approximation can promote development of a more elaborated
many-electron non-stationary theory of {\sc ac}-driven graphene
nanostructures which is crucial for future graphene-based
electronics.

SS acknowledges support from the Alexander von Humboldt foundation
through the Bessel prize and thanks Sasha Alexandrov and Viktor
Kabanov for stimulating discussions. PH thanks for support by the
cluster of excellence, Nanosystems Initiative Munich (NIM).


\begin{references}
\bibitem{reviews} K.S. Novoselov {\it et al.,}
Nature {\bf 438}, 197 (2005);
A.H. Castro Neto {\it et al.,}
Rev.\ Mod.\ Phys.\ {\bf 81}, 109 (2009);
A.V. Rozhkov {\it et al.,}
Phys.\ Reports {\bf 503}, 77 (2011). 
\bibitem{paradox} M.I. Katsnelson, K.S. Novoselov, A.K. Geim, Nature Phys. {\bf 2}, 620 (2006).
\bibitem{paradox-exp} N. Stander, B. Huard, D. Goldhaber-Gordon, Phys.\ Rev.\ Lett.\ {\bf 102}, 026807 (2009);
A.F. Young, P. Kim, Nature Phys.\ {\bf 5}, 222 (2009);
S.-G. Nam {\it et al.,}
Nanotechnology {\bf 22}, 415203 (2011).
\bibitem{superl} C.X. Bai, X.D. Zhang, Phys.\ Rev.\ B {\bf 76} 075430 (2007);
C.H. Park {\it et al.,}
Nature Physics {\bf 4}, 213 (2008);
C.H. Park {\it et al.,}
Phys. Rev. Lett. {\bf 101}, 126804 (2008);
M. Barbier, P. Vasilopoulos, F.M. Peeters, Phys. Rev. B {\bf 81}, 075438 (2010);
L.Z. Tan, C.H. Park, S.G. Louie, Phys. Rev. B {\bf 81}, 195426 (2010).
\bibitem{super5} Y.P. Bliokh {\it et al.,}
Phys.\ Rev.\ B {\bf 79} 075123 (2009).
\bibitem{left} V.V. Cheianov, V.I. Fal'ko, B.L. Altshuler, Science {\bf 315}, 1252 (2007);
V.A. Yampol'skii, S. Savel'ev, F. Nori, New J. Phys. {\bf 10}, 053024 (2008).
\bibitem{peeters09b} M. Barbier, P. Vasilopoulos, F.M. Peeters, Phys. Rev B {\bf 80}, 205415 (2009).
\bibitem{pnexperiments} H.-Y. Chiu {\it et al.,}
Nano Lett. {\bf 10}, 4634 (2010);
M.Y. Han {\it et al.,}
Phys. Rev. Lett. {\bf 98}, 206805 (2007);
B. Huard {\it et al.,}
Phys. Rev. Lett. {\bf 98}, 236803 (2007).
\bibitem{magstruc} T.K. Ghosh {\it et al.,} Phys.\ Rev.\ B {\bf 77}, 081404(R) (2008);
W. H\"ausler {\it et al.,} Phys.\ Rev.\ B {\bf 78}, 165402 (2008);
W. H\"ausler, R. Egger, Phys.\ Rev.\ B {\bf 80}, 161402(R) (2009).
\bibitem{Efetov} M.V. Fistul, K.B. Efetov, Phys. Rev. Lett. {\bf 98}, 256803 (2007).
\bibitem{laser1} H.L. Calvo {\it et al.,}
Appl. Phys. Lett. {\bf 98}, 232103 (2011).
\bibitem{laser2} S.E. Savel'ev, A.S. Alexandrov, Phys. Rev. B {\bf 84}, 035428 (2011).
\bibitem{kanemele} C.L. Kane, E.J. Mele, Phys. Rev. Lett. {\bf 95}, 226801 (2005).
\bibitem{zigzagribbons} A. Rycerz, J. Tworzydlo, C.W.J. Beenakker, Nature Physics
{\bf 3}, 172 (2007);
A.R. Akhmerov {\it et al.,}
Phys. Rev. B {\bf 77}, 205416 (2008);
J.M. Pereira {\it et al.,}
J. Phys.: Condens. Matter {\bf 21}, 045301 (2009).
\bibitem{garcia08} J.L. Garcia-Pomar, A. Cortijo, M. Nieto-Vesperinas, Phys. Rev. Lett. {\bf 100}, 236801 (2008).
\bibitem{ziegler10} A. Hill, A. Sinner, K. Ziegler, New J. Phys. {\bf 13}, 035023 (2011).
\bibitem{ando98} T. Ando, T. Nakanishi, R. Saito, J. Phys. Soc. Jpn. {\bf 67}, 2857 (1998).
\bibitem{zhou06} S. Y. Zhou {\it et al.,}
Nature Physics {\bf 2}, 595 (2006).
\bibitem{wkb-method} V.V. Cheianov, V.I. Fal'ko,
Phys.\ Rev.\ B {\bf 74}, 041403(R) (2006);
P.G. Silvestrov, K.B. Efetov, Phys. Rev. Lett. {\bf 98}, 016802 (2007);
S.V. Syzranov, M.V. Fistul, K.B. Efetov, Phys. Rev. B {\bf 78}, 045407 (2008).
\bibitem{harmonicgates} B. Trauzettel, Ya.M. Blanter, A.F. Morpurgo,
Phys. Rev. B {\bf 75}, 035305 (2007); P. San-Jose {\it et al.,}
Phys. Rev. B {\bf 84}, 155408 (2011).
\bibitem{solomon} D. Solomon, Can. J. Phys. {\bf 88}, 137 (2010).
\bibitem{couranthilbert} R. Courant, D. Hilbert, {\it Methods of Mathematical Physics}, Volume II, Wiley-Interscience (1962).
\bibitem{tinkham}M. Tinkham, {\it Introduction to Superconductivity}, Dover Publications Inc. (2004).
\bibitem{woods} H. Raether, {\it Surface Plasmons}, Springer, New York (1988).
\end{references}
\end{document}